\documentclass[12pt]{article}
\usepackage{amssymb,amsmath,amscd,amsthm}

\newtheorem{theorem}{Theorem}[section]

\newtheorem{lemma}[theorem]{Lemma}

\font\bbc=msbm10 scaled 1200

\newcommand{\R}{\mbox {\bbc R}}

\date{}

\begin{document}

\title{On critical value of the coupling constant in exterior elliptic problems.}

 \author{Rajan Puri\thanks{Department
of Mathematics and Statistics, University of North Carolina,
Charlotte, NC 28223, USA, (rpuri7@uncc.edu). } \and
 Boris Vainberg\thanks{Department
of Mathematics and Statistics, University of North Carolina,
Charlotte, NC 28223, USA, (brvainbe@uncc.edu). The second author was supported  by the NSF grant DMS-1714402 and the Simons Foundation grant 527180. Corresponding author}}

\date{}

%\begin{document}

\maketitle

\begin{abstract}
We consider exterior elliptic problems with coefficients stabilizing at infinity and study the critical value $\beta_{cr}$ of the coupling constant (the coefficient at the potential) that separates operators with a discrete spectrum and those without it. The dependence of $\beta_{cr}$ on the boundary condition and on the distance between the boundary and the support of the potential is described. The discrete spectrum of a non-symmetric operator with the FKW boundary condition (that appears in diffusion processes with traps)  is also investigated.
\end{abstract}



\section{Introduction} The paper concerns the negative spectrum of the following exterior problem
\begin{equation}\label{1}
-{\rm div} ( a(x)\nabla u ) -\beta V(x)u=\lambda u, \quad x\in \Omega\subset R^d; \quad Bu =0, ~~x\in \partial \Omega;~~~\beta\geq 0,
\end{equation}
where $\Omega$ is the exterior of a bounded domain with a smooth boundary, $a$ is a smooth function such that $a(x)=1$ when $|x|>R$, the potential $V$ is nonnegative and equal to zero when $|x|>R$, and operator $B$ in the boundary condition stays for the Dirichlet $u|_{\partial\Omega}=0$ or Neumann $\frac{\partial u}{\partial n}|_{\partial\Omega}=0$ boundary condition.

The operator
\begin{equation}\label{H0}
H_0:=-{\rm div} (a(x)\nabla) :L^2(\Omega)\to L^2(\Omega),
\end{equation}
defined on the space of Sobolev functions $H^2(\Omega)$ satisfying the Dirichlet or Neumann boundary condition is non-negative, and its spectrum is a.c. and coincides with the positive semi-axis $[0,\infty)$. The relatively compact negative perturbation by the potential term in (\ref{1}) can produce a discrete negative spectrum
$\{\lambda_j \}$.  Since the operator $H_\beta=H_0-\beta V(x)$ is bounded from below, we have  $\lambda_j\geq-\beta\max V(x)$. It is also proved in \cite[Theorem 10]{75} that the negative eigenvalues are separated from the origin. Thus the discrete spectrum consists of at most finitely many negative eigenvalues $\lambda_j, 1\leq j\leq N$. Alternatively, the latter conclusion can be obtained from the Cwikel-Lieb-Rozenblum inequality \cite{cw,li,ro}:
\begin{equation}\label{clr}
\#\{\lambda_j<0\}\leq C_d\int_{\partial\Omega}(\beta V)^{d/2}dx,
\end{equation}
which is valid when $d\geq 3$, or from the Bargmann type estimate \cite{barg} when $d=1,2$ (the latter case requires some preliminary work on estimation of the heat kernel $p(t,x,x),~t\to\infty,$ that corresponds to the operator $H_0-\varepsilon q_1(x)$ with a small killing potential $\varepsilon q_1(x)\geq 0$).

The main question under investigation is whether the discrete spectrum appears for arbitrarily small perturbations (arbitrarily small $\beta>0$), or $\beta$ must be large enough to create negative eigenvalues. Denote by $\beta_{cr}$ the value of $\beta$ such that operator $H_\beta=H_0-\beta V(x)$ does not have negative eigenvalues for $\beta<\beta_{cr}$ and has them if $\beta>\beta_{cr}$.
 Thus we would like to know when $\beta_{cr}=0$ and when $\beta_{cr}>0$. The answer is known \cite{simon}, \cite{pol} for the Schr\"{o}dinger operator
$-\Delta-\beta V(x)$ in $R^d$ and depends only on dimension: $\beta_{cr}=0$ if $d=1,2$ and $\beta_{cr}>0$ if $d\geq 3$. The value of $\beta_{cr}$ remains positive for operator (\ref{1}) with both Dirichlet and Neumann boundary conditions if $d\geq 3$. This fact follows immediately from (\ref{clr}). Indeed, this
inequality implies that the negative eigenvalues do not exist if $\beta$ is so small that the right-hand side in (\ref{clr}) is less than one. If $d=1$ or $ 2$, then
(\ref{clr}) is not applicable, and the answer to the main question for the problem (\ref{1}) is different from the answer in the case of the Schr\"{o}dinger operator. In fact, it depends on the boundary condition.

The main result will be based on the following statement proved in the next section.
%The plan of the paper is as follows.
Consider the truncated resolvent $A_\lambda$  of operator $H_0$ with the cut-off function $\chi=\sqrt{V(x)}$:
\begin{equation}\label{A0}
A_\lambda=\sqrt{V}(H_0-\lambda)^{-1}\sqrt{V}:L^2(\Omega)\to L^2(\Omega), ~~\lambda<0.
\end{equation}
It will be shown that the choice $\beta_{cr}>0$ or $\beta_{cr}=0$  depends on whether the truncated resolvent $A_\lambda$ is bounded or goes to infinity when $\lambda\to 0^-$. In fact, $\beta_{cr}$ will be expressed through $\|A_\lambda\|$. Note that operator $H_\beta-\lambda$ decays when $\lambda$ grows, and therefore $\|A_\lambda\|$ is monotone in $\lambda$ and the limit $\lim_{\lambda\to0^-}[1/\|A_\lambda\|]$ exists.
\begin{theorem}\label{tres}
We have $\beta_{cr}=1/\|A_{0^-}\|$. Thus
if $\|A_\lambda\|<C<\infty$ as $\lambda\to 0^-$, then $\beta_{cr}>0$; if $\|A_\lambda\|\to\infty$ as $\lambda\to 0^-$, then $\beta_{cr}=0$.
\end{theorem}
Next, we will prove
\begin{theorem}\label{tmr}
Let $d=1$ or $ 2$. Then $\|A_{0^-}\|<\infty$ and $\beta_{cr}>0$ in the case of the Dirichlet boundary condition, and $\|A_{0^-}\|=\infty$ and $\beta_{cr}=0$ in the case of the Neumann boundary condition.
\end{theorem}
{\bf Remark.} The boundedness/unboundedness of $A_\lambda$ at $\lambda\to 0^-$ is a manifestation of the transiency/recurrency of the random walk with the generator $H_0$.
% A direct analytical approach to obtain asymptotic expansion of $A_\lambda, \lambda\to 0,$ for general elliptic operators of higher order can be found in \cite{75} (see Theorems...).  Modified proofs (that complete the proof of  Theorem \ref{tmr}) will be given in section 3 of the present paper.

Section 3 is devoted to the study of the dependence of $\beta_{cr}$ (when it is positive) on the distance of the support of the potential $V$ from the boundary of the domain.

In the last section, the discrete spectrum is studied for exterior boundary problems with a non-standard boundary condition of the form
\begin{equation}\label{fk1}
-{\rm div} (a(x)\nabla u) -\beta V(x)u-\lambda u=f, \quad x\in \Omega\subset R^d; \quad
u|_{\partial\Omega}=\alpha ,\quad  \int_{\partial\Omega} \frac{\partial u}{\partial n}d \mu=0,
\end{equation}
where $\alpha$ is an arbitrary constant and $\mu$ is a probability measure on $\partial\Omega$. The latter boundary condition  will be called FKW-condition by the first letters of the authors' names (Freidlin, Koralov, Wentzell) who introduced it recently in \cite{fkw},  \cite{fkw1} in a slightly different setting. This FKW condition appears in the description of the diffusion process in $\Omega$ that is a limit, as $A\to\infty$, of the process in $R^d$ with a large drift $A\overrightarrow{F}(x)\cdot\nabla u$ in $R^d\backslash\Omega$, where the vector field $\overrightarrow{F}$ is directed to an interior point of $R^d\backslash\Omega$, and the time spent by the process outside of $\Omega$ is not taken into account. For simplicity, we will assume in the last section that $\partial\Omega $ and $a(x)$ are infinitely smooth. The results below can be easily extended to a more general situation (which is considered in \cite{fkw}, \cite{fkw1}) when $R^d\backslash\Omega$ is a union of several non-intersecting domains, and FKW conditions (with different $\alpha, \mu$) are imposed on the boundaries of these domains.

We will prove that the spectrum of problem (\ref{fk1}) consists of the continuous component $[0,\infty)$ and a discrete set of eigenvalues with the only possible limiting point at infinity. It will be shown that $\|A_{0^-}\|<\infty$ for problem  (\ref{fk1}) if $d\geq 3$ and $\|A_{0^-}\|=\infty$ if $d=2$. Problem (\ref{fk1}) is not symmetric and may have complex eigenvalues. Moreover, eigenvalues can be imbedded into the continuous spectrum (compare with \cite{klv}). If $\mu$ is the Lebesgue measure on the boundary, then the problem is symmetric and may have only real eigenvalues $\lambda\leq 0$. In the latter case, Theorem  \ref{tres} and its proof remain valid, and therefore $\beta_{cr}>0$ for problem  (\ref{fk1}) when $d>2$, $\beta_{cr}=0$ when $d=1$ or $2$.

% The arguments below could be made a little simpler if we assume that $\mu$ (see (\ref{fk})) belongs to the Sobolev space $H^{-1/2}(\partial\Omega)$. Then the problem (\ref{1}), (\ref{fk}) will be well defined for $u\in H^2(\Omega)$. However, it is important for applications to consider arbitrary measures $\mu$, for example delta functions on the boundary. Thus we will use Shauder estimates for solutions of (\ref{1}), (\ref{fk}), i.e., we will assume that $u\in C^{2,\alpha}$, and the boundary of the domain and coefficients of the equation are smooth enough to guarantee the validity of the Shauder estimetes in the space  $C^{2,\alpha}$.

\section{Proof of Theorems \ref{tres}, \ref{tmr}.}

We will need the following lemma.
\begin{lemma}\label{lal}
There is a one-to-one correspondence between the eigenspaces of operators $H_\beta$, $\beta> 0$,  and $A_\lambda, ~\lambda<0$. Namely, if $u\in H^2(\Omega)$ is an eigenfunction of $H_\beta$ with an eigenvalue $\lambda<0$, then $w=\sqrt Vu$ is an eigenfunction of operator $A_\lambda$ with the eigenvalue $\frac{1}{\beta}.$ Vice versa, if $w\in L^2(\Omega)$ is an eigenfunction of $A_\lambda, ~\lambda<0,$ with an eigenvalue $\mu$, then $\mu>0$ and $u=(H_0-\lambda)^{-1}(\sqrt{V}w)$ is an eigenfunction of $H_{1/\mu}$ with the eigenvalue $\lambda$.
\end{lemma}
{\bf Proof.} Let $H_\beta u=\lambda u$. Then $(H_0-\lambda )u=\beta V u$ and $u=\beta (H_0-\lambda)^{-1}(V u)$. After multiplying both sides by $\sqrt V$, we obtain $w=\beta A_\lambda w$, i.e., $w $ is an eigenfunction of $A_\lambda$ with the eigenvalue $1/\beta$.

Conversely, let $A_\lambda w=\mu w, ~\lambda<0$, i.e.,
\begin{equation}\label{13}
\sqrt V(H_0-\lambda)^{-1}(\sqrt V w)=\mu w.
\end{equation}
Since operator $A_\lambda, ~\lambda<0,$ is positive, we have $\mu>0$. Define $u=(H_0-\lambda)^{-1}(\sqrt V w)$. Then $u\in H^2, Bu|_{\partial\Omega}=0,$ and $(H_0-\lambda )u=\sqrt V w$. We multiply both sides of (\ref{13}) by $\sqrt V$ and express $\sqrt Vw$ through $u$. This leads to $Vu=\mu(H_0-\lambda )u$, i.e., $u$ is an eigenfunction of $H_{1/\mu}$ with the eigenvalue $\lambda$.

\qed

{\bf Proof of Theorem \ref{tres}.} Operator $(H_0-\lambda)^{-1}:L^2(\Omega)\to H^2(\Omega),~\lambda<0,$ (where $H^2(\Omega)$ is the Sobolev space) is bounded and $V$ has a compact support. Hence, Sobolev's imbedding theorem implies that operator (\ref{A0}) is compact. Since operator  $A_\lambda,~ \lambda<0,$ is positive, depends continuously on $\lambda$, and increases when $\lambda$ increases, its principal (largest) eigenvalue $\mu_0(\lambda), ~\lambda<0,$ is a positive, continuous,  and monotonically increasing function of $\lambda$. Let $\mu^*=\lim_{\lambda\to 0^-}\mu_0(\lambda)=\|A_{0^-}\|$. Obviously, $\|A_\lambda\|\to 0$ as $\lambda\to -\infty$. Thus, the range of the function $\mu_0(\lambda), ~-\infty<\lambda<0, $ is $(0,\mu^*)$. Hence, for each $\mu\in(0,\mu^*)$, there is a $\lambda=\lambda_0<0$ such that $\mu_0(\lambda_0)=\mu$, and therefore $H_{1/\mu}$ has the eigenvalue $\lambda=\lambda_0$ due to Lemma \ref{lal}. Since $1/\mu\in (1/\mu^*,\infty)$, operator $H_{\beta}$ has at least one negative eigenvalue when $\beta>1/\mu^*$.

Since $A_\lambda,~\lambda<0,$ can not have eigenvalues larger than $\mu^*$, Lemma  \ref{lal} implies that $H_\beta$ does not have eigenvalues $\lambda<0$ if $\beta<1/\mu^*$. It remains to recall that $\mu^*=\|A_{0^-}\|$.

\qed

In order to prove Theorem \ref{tmr}, we will need the following lemma.
\begin{lemma}\label{lls}
Let $\omega \subset \R^d$ be a bounded domain with a Lipschitz boundary. Let H be a non-zero closed subspace of Sobolev space $H^1(\omega)$ that
does not contain non-zero constant functions. Then there exists a constant $ C>0$ that depends on $\omega$ such that
\begin{equation}\label{puan}
    \|u\|_{L^2} \leq C\|\nabla u\|_{L^2}, \quad u\in H.
\end{equation}
\end{lemma}
{\bf Proof.}
Assume that (\ref{puan}) is not true. Then there is a sequence $v_n\in H,~ n\in\mathbb N,$ such that
$\|v_n\|_{L^2} \geq n \|\nabla v_n\|_{L^2}$.
Define $u_n= \frac{v_n}{\|v_n\|_{L^2}}$. Then $ \|u_n\|=1$ and
$\|\nabla u_n\|_{L^2}\leq \frac{1}{n}$, i.e.,  $\{u_n\}$ is a bounded sequence in $H^1(\omega)$. Since the imbedding $H^1(\omega) \subset L^2(\omega)$ is compact, there exists a subsequence of $\{ u_n\}$ that converges in $L^2(\omega)$. Without loss of generality we can assume that $\{ u_n\}$ converges in $L^2(\omega)$ as $n\to\infty$. Since $\|\nabla u_n\|_{L^2}\to 0$ as $n\to\infty$, the sequence $\{ u_n\}$ converges in $H^1(\omega)$. The limiting function $u$ belongs to $H$ since $u_n\in H$ and $H$ is a closed subspace of $H^1(\omega)$. Relation $\|\nabla u_n\|_{L^2}\to 0$ as $n\to\infty$ implies that $u$ is a constant. This constant must be zero since $H$ does not contain non-zero constant functions. The latter contradicts the fact that $\|u_n\|_{L^2}=1$. Hence, our assumption is wrong.

\qed

{\bf Proof of Theorem \ref{tmr}.} Consider first the case of the Dirichlet boundary condition. We would like to show that $\|A_\lambda\|<C<\infty, ~\lambda\to 0^-$. From (\ref{A0}) it follows that $A_\lambda f=\sqrt Vu$, where $u=(H_0-\lambda)^{-1}\sqrt Vf$, i.e., $(H_0-\lambda)u=\sqrt Vf$. From the Green formula it follows that
\[
\int _\Omega (a(x)|\nabla u|^2-\lambda |u|^2)dx=\int_{\Omega_R} \sqrt Vfudx.
\]
Hence,
\[
\int _\Omega |\nabla u|^2dx\leq\int_{\Omega_R} |\sqrt Vfu|dx, \quad \lambda<0.
\]
Lemma \ref{lls} implies that
\[
\int _{\Omega_R} | u|^2dx\leq C\int _\Omega |\nabla u|^2dx\leq C\int_{\Omega_R} |\sqrt Vfu|dx\leq
\frac{1}{2}\int_{\Omega_R} |u|^2dx+\frac{C^2}{2}\int_{\Omega_R} |\sqrt Vf|^2dx, \quad \lambda<0.
\]
Thus
\[
\|u\|_{L^2(\Omega_R)}\leq C\|\sqrt Vf\|_{L^2(\Omega_R)}\leq C_1\|f\|_{L^2(\Omega_R)}, \quad \lambda<0,
 \]
and
 \[
 \|A_{\lambda}f\|_{L^2(\Omega)}=\|\sqrt Vu\|_{L^2(\Omega_R)}\leq C_2\|f\|_{L^2(\Omega)}, \quad \lambda<0,
  \]
Hence, $\|A_\lambda\|\leq C, \quad \lambda\to0^-$, and Theorem \ref{tres} implies that $\beta_{cr}>0$.

Consider now operator (\ref{H0}) with the Neumann boundary condition in dimensions $d=1$ and $2$. It was shown in \cite{pol} that the Schr\"{o}dinger operator $H=c\Delta-\beta V$ in $R^d,~d=1,2$, with arbitrary constants $c,\beta>0$ has negative eigenvalues. Let $\psi$ be its eigenfunction with an eigenvalue $\lambda<0$. Then
\[
<H\psi,\psi>=\int_{R^d}(c|\nabla\psi|^2-\beta V|\psi|^2)dx=\lambda\|\psi\|^2<0.
\]
We choose $c=\max a(x)$. Since the support of $V$ belongs to $\overline{\Omega}$, we have
\[
<H_\beta\psi,\psi>=\int_{\Omega}(a(x)|\nabla\psi|^2-\beta V|\psi|^2)dx\leq \int_{R^d}(c|\nabla\psi|^2-\beta V|\psi|^2)dx<0.
\]
Thus $H_\beta$, with an arbitrary $\beta>0$, has negative eigenvalues. Hence $\beta_{cr}=0$ (and, therefore, $\|A_{0^-}\|=\infty$.)

\qed

%For convenience of readers, we will provide two proofs of the letter fact: a short proof based on the Kaz-Feinman formula and the recurrence %arguments, and a much longer proof for those who prefer to avoid the probabilistic approach.

\section{Potentials with the supports near the boundary.}
This section is devoted to the dependence of $\beta_{cr}$ on the distance between the support of the potential and the boundary of the domain. In fact, it is obvious that moving the support of the potential to the boundary does not affect  $\beta_{cr}$ essentially, but we will shrink the size of the support, increase the height of the potential appropriately (see below), and move the potential toward to the boundary. It will be shown that this process, in the case of the Dirichlet boundary condition, will imply the blowing up of  $\beta_{cr}$ in dimension $d=1$. In dimension two (and the Dirichlet boundary condition), the behavior of $\beta_{cr}$ depends on the relation between the rates of the shrinking of the support of the potential and the rate of its motion to the boundary. We do not consider the Neumann boundary condition when $d=1$ or $2$ since $\beta_{cr}$ is always zero in this case. We will show that $\beta_{cr}$ is not very sensitive  to the location of the support of the potential for both Dirichlet and Neumann problems if $d\geq 3$.

For the sake of the transparency, we will assume that $a(x) \equiv 1$, i.e., problem (\ref{1}) has the form
\begin{equation}\label{11a}
-\Delta u -\beta V_n(x)u=\lambda u, \quad x\in \Omega\subset R^d; \quad u|_{\partial \Omega} =0;~~~\beta\geq 0, \quad n\to\infty.
\end{equation}
We will consider the potential $V_n$ of the form $V_n(x)=h_d(n)W((x-x(n))n)$, where $W\in C_0(R^d),~W\geq0,$ the support of $W$ belongs to the unit ball, $x(n)\to x_0\in \partial\Omega$ as $n\to\infty$, the support of $V_n(x)$ belongs to $\Omega$, and $h_d(n)$ will be chosen in the next paragraph.

Let $d\geq 3$, so that $\beta_{cr}>0$. In order to study the dependence of $\beta_{cr}$ on the location of the potential, we consider the problem in the whole space $R^d$, assume that $x(n)=0$, and choose $h_d(n)$ in such a way that $\beta_{cr}$ does not depend on $n$. This value of $h_d(n)$ will be used in (\ref{11a}) to study the dependence of $\beta_{cr}$ on the location of the potential. We will proceed similarly when $d=1,2$. By Theorem \ref{tres}, $h_d(n)$, with $d\geq 3$, must be chosen in such a way that the norm of the operator $A_{0^-}=A_{0^-}(n)$ with the integral kernel
$$
A_{0^-}(x,y,n)=\sqrt{h_dW(xn)}\frac{c_d}{|x-y|^{d-2}}\sqrt{h_dW(yn)}
$$
(where $c_d$ is a constant) does not depend on $n$. The substitution $xn=x',~yn=y'$ implies that $h_d(n)=n^{2}, ~d\geq 3$. Indeed, this substitution immediately implies that if $u(x)$ is an eigenfunction of the operator $A_{0^-}(n)$ with an eigenvalue $\lambda(n)$, then $u(x'/n)$ is an eigenfunction of the operator $A_{0^-}(1)$ with the eigenvalue $\lambda(1)=\frac{h_d(1)n^2}{h_d(n)}\lambda(n)$. The converse relation is also valid. Hence, the choice $h_d(n)=n^{2}, ~d\geq 3,$ implies that $\|A_{0^-}(n)\|$ does not depend on $n$.

A small change is needed in dimensions one and two.  We can't consider the operator $A_{0^-}(n)$ in the whole space for small dimensions (the operator is not defined), but we can consider a similar operator for the Dirichlet problem in $\Omega$. Its integral kernel is bounded when $d=1$ and has the singularity $\frac{1}{2\pi}\ln\frac{1}{|x-y|}$ if $d=2$. The same substitution implies that $h_1(n)=n,~h_2(n)=\frac{n^2}{\ln n}$. The norm of $A_{0^-} (n)$ depends on $n$ in this case, but approaches a constant as $n\to\infty$.

For transparency, we will not study $\beta_{cr}$ in the general setting, but focus our attention on the case when $\partial\Omega$ contains a flat part $\Gamma,~x_0$ is an interior point of $\Gamma$, and $x(n)$ moves toward $x_0$ in the direction perpendicular to $\Gamma$.
\begin{theorem}\label{tlast}
If $d=1$, then $\beta_{cr}$ (for operator (\ref{11a})) goes to infinity as $n\to\infty$. The same is true if $d=2$ and $|x(n)-x_0|<C/n,~n\to\infty$. If $d=2$ and $|x(n)-x_0|\to 0,~|x(n)-x_0|>C/n^\delta,~n\to\infty,$ with some $\delta\in(0,1)$, then $\beta_{cr}$ remains bounded as $n\to\infty$. If $d\geq 3$, then $\beta_{cr}$ remains bounded as $n\to\infty$ for both the Dirichlet and Neumann boundary conditions.
\end{theorem}
{\bf Remarks.} The arguments in the proof allow one to estimate the rate with which $\beta_{cr}$ tends to infinity. This rate depends on the rate of the convergence of $x(n)$ to $x_0$.

{\bf Proof.}  Let $d=1$. Since the exterior of an interval is a union of two half-lines, it is enough to prove the statement for the Dirichlet problem on $(0,\infty)$. The Green function $G_\lambda$ for the operator
\[
H_0u=-u''-\lambda u, \quad x>0, \quad u(0)=0, \quad \lambda<0,
\]
has the form $G_\lambda=\frac{e^{-k|x-\xi|}-e^{-k|x+\xi|}}{-2k}, ~x,\xi>0,~ k=\sqrt{|\lambda|},$ and its limiting value as $\lambda\to0^-$ is $|x+\xi|-|x-\xi|$. Hence, the operator $A_{0^-}(n)$ defined by (\ref{A0}) has the integral kernel
\[
A_{0^-}(x,\xi,n)=n\sqrt{W((x-x(n))n)}(|x+\xi|-|x-\xi|)\sqrt{W((\xi-x(n))n)}, \quad x,\xi>0.
\]
Since $|x-x(n)|+|\xi-x(n)|<\frac{c}{n}$ on the support of $A_{0^-}(x,\xi,n)$, and $x(n)\to0, ~n\to\infty$, there exists $\alpha(n)$ such that $\alpha(n)\to 0, ~n\to\infty,$ and
\[
|x+\xi|-|x-\xi|\leq \alpha(n)
\]
on the support of $A_{0^-}(x,\xi,n)$. Then one can easily see that
\[
\|A_{0^-}(n)\|\leq [\int_0^\infty \int_0^\infty A_{0^-}^2(x,\xi,n)dxd\xi]^{1/2}\leq C\alpha(n)\to 0,~~~n\to\infty,
\]
and the statement of Theorem \ref{tlast} for $d=1$ follows from Theorem \ref{tres}.

Let us consider the case $d\geq 3$. Without loss of generality, we can assume that $\Gamma$ is a part of the hyperplane $x_1=0,~x_0=0,$ and there exists a ball $B_\varepsilon$ of radius $\varepsilon$ centered at the origin such that its right half $B_\varepsilon^+$, where $x_1>0$, belongs to $\Omega$, and the other half does not contain points of $\Omega$. Hence $x(n)$ moves to the origin along the positive $x_1$-semi-axis as $n\to\infty$. Let $E(x)=\frac{c_d}{|x|^{d-2}}$ be a fundamental solution of $-\Delta$. For $\xi\in B_\varepsilon$, denote by $\xi^*$ the point symmetrical to $\xi$ with respect to the plane $x_1=0$.
\begin{lemma}\label{lll}
The Green function $G=G_\mp(x,\xi)$ of the Dirichlet (Neumann) problem in $\Omega $ for the operator $-\Delta$ has the form $G_\mp=E(x-\xi)\mp E(x-\xi^*)+F(x,\xi)$, where $F$ is uniformly bounded when $x\in \Omega,~\xi\in B_{\varepsilon/2}$.
\end{lemma}
{\bf Remark.} Additional smoothness of $\partial\Omega$ is needed to prove this statement in the case of the Neumann boundary condition. For example, one can assume that $\partial\Omega\in C^{2,\alpha}$.

{\bf Proof.} In the case of the Dirichlet problem, $F$ is the solution of the homogeneous equation $\Delta F=0$ with the boundary condition
\[
F=E(x-\xi^*)-E(x-\xi), \quad x\in\partial\Omega,~~\xi\in B_{\varepsilon/2}.
\]
 Since $F|_{\partial\Omega}$ is bounded uniformly in $\xi\in B_{\varepsilon/2}$, the maximum principle implies that $|F|<C$. In the case of the Neumann boundary condition, the normal derivative of $F$ on the boundary belongs to $C^{1,\alpha}$ (if $\partial\Omega\in C^{2,\alpha}$). From local a priori estimates for elliptic equations, it follows that $F|_{\partial\Omega}\in C^\alpha$, and the maximum principle can be applied again.

\qed

In order to  prove the theorem in the case $d\geq 3$, it is enough to show that $\|A_{0^-}(n)\|\geq c>0$ when $n\to \infty$ (see Theorem \ref{tres}). Let $\widehat{F}$ be the operator in $L^2(\Omega)$ with the integral kernel
$$
\widehat{F}(x,\xi)=\sqrt{n^2W((x-x(n))n)}F(x,\xi)\sqrt{n^2W((\xi-x(n))n)},
$$
where $F$ is defined in lemma above. The support of $\sqrt{n^2W((\xi-x(n))n)}$ belongs to $B_{\varepsilon/2}$ when $n$ is large enough, and therefore Lemma \ref{lll} and the substitution
\begin{equation}\label{subst}
x-x(n)=y/n,~\xi-x(n)=\sigma/n
\end{equation}
imply that
 $$
 \int_\Omega
\int_\Omega \widehat{F}^2(x,\xi)dxd\xi\leq\frac{C}{n}\to 0, \quad n\to\infty.
$$
Hence, $\|\widehat{F}\|\to 0,~n\to\infty$, and it remains to show that the norm of the operators $\widehat{E}_\mp$ in $L^2(\Omega)$ with the integral kernel
$$
c_dn^2\sqrt{W((x-x(n))n)}[|x-\xi|^{2-d}\mp |x-\xi^*|^{2-d}]\sqrt{W((\xi-x(n))n)}
$$
 is bounded from below when $n\to\infty$. One can consider these operators in $L^2(R^d_+),~R^d_+=\{x:x_1>0\}$ instead of $L^2(\Omega)$ since the integral kernel vanishes if $x$ or $\xi$ are not in $B^+_{\varepsilon/2}$ and $n$ is large enough. The norm remains the same after substitution (\ref{subst}) (since the principal eigenvalues are the same). Hence, it is enough to show that the norm of the integral operator $G_\mp$ in $L^2(R^d_+)$ with the integral kernel
\[
c_d\sqrt{W(y)}[|y-\sigma|^{2-d}\mp| y-\sigma^*+2nx(n)|^{2-d}]\sqrt{W(\sigma)}
\]
 is bounded from below when $n\to\infty$.

Consider an arbitrary ball $B\in R^d_+$ such that its distance from the origin is positive and $W(x)\geq \alpha>0$ when $x\in B$. There exists $\rho>0$ such that $| y-\sigma^*+2nx(n)|^{2-d}\leq (1-\rho)|y-\sigma|^{2-d},~y,\sigma\in B$, i.e.,
\[
|y-\sigma|^{2-d}\mp| y-\sigma^*+2nx(n)|^{2-d}\geq \rho |y-\sigma|^{2-d}.
\]
Hence $\|G_\mp\|$ is not smaller than the norm of the  operator in $L^2(B)$ with the integral kernel $c_d\alpha\rho|y-\sigma|^{2-d}$, which does not depend on $n$. This completes the proof of the theorem in the case of $d\geq 3$.

The Dirichlet problem when $d=2$ is treated absolutely similarly to the case $d\geq3$. The only difference is that operator $G_-$ now has the following integral kernel:
\begin{equation}\label{12121}
\frac{1}{2\pi\ln n}\sqrt{W(y)}\ln\frac{| y-\sigma^*+2nx(n)|}{|y-\sigma|}\sqrt{W(\sigma)}.
\end{equation}
If $n|x(n)|<C$, then operator $G_-$ converges strongly to zero as $n\to \infty$. Hence, $A_{0^-}(n)$ has the same property, and $\beta_{cr}\to \infty$ due to  Theorem \ref{tres}. If $|x(n)|>Cn^{-\delta},~0<\delta<1,$ then we write the logarithm of the quotient in (\ref{12121}) as the difference of the logarithms and represent the operator $G_-$as $G_-=G_1-G_2$. Obviously, $G_2$ converges strongly to zero as $n\to \infty$, and $G_-$ is bounded from below for large $n$ by the operator with the kernel
$$
\frac{\ln| 2nx(n)|}{2\pi\ln n}\sqrt{W(y)}\sqrt{W(\sigma)}\geq \frac{1-\delta}{4\pi}\sqrt{W(y)}\sqrt{W(\sigma)}.
$$
Hence, $\|A_{0^-}(n)\|$ is bounded from below as $n\to\infty$, and $\beta_{cr}$ is bounded.

\qed
\section{FKW exterior boundary problem}
Recall that in this section we assume, for simplicity, that $\partial\Omega$ and $a(x)$ are infinitely smooth. We will study problem (\ref{fk1}) with $f\in H^s(\Omega)$ and $u\in H^{s+2}(\Omega)$, where $s>[\frac{d}{2}]$. The last restriction and the Sobolev imbedding theorem imply the inclusion $u\in C^1({\overline{\Omega}})$, which  makes the last condition in (\ref{fk1}) meaningful.

We will use the same notation $H_0$ for the operator related to problem (\ref{fk1}):
\[
H_0:H^{s+2}(\Omega)\to H^s(\Omega),
\]
where the domain of $H_0$ consists of functions $u\in H^{s+2}(\Omega) $ satisfying the last two boundary conditions in (\ref{fk1}). Obviously, $[0,\infty)$ belongs to the continuous spectrum of $H_\beta=H_0-\beta V(x)$ since one can use the same Weyl sequence for operator $H_\beta-\lambda, \lambda>0,$ as the one in the case of the Dirichlet or Neumann boundary conditions. To study the spectrum outside of $[0,\infty)$ (and eigenvalues on $[0,\infty)$ ), consider the resolvent
$R_\lambda=(H_\beta-\lambda)^{-1}$ and the truncated resolvent $\widehat{R}_\lambda=\chi(x)R_\lambda\chi(x)$, where $\chi\in C^\infty_0$.
\begin{theorem}\label{t41}
1) The resolvent $R_\lambda$ is meromorphic in $\lambda\in \mathbb C\backslash [0,\infty)$. Its poles do not have limiting points except, possibly, at infinity.

2) If $k=\sqrt \lambda, {\rm Im}k>0, $ then the truncated resolvent $\widehat{R}_{k^2},~{\rm Im}k>0, $ has a meromorphic continuation to the whole complex $k$-plane when $d$ is odd or to the Riemann surface of $\ln k$ when $d$ is even. The poles in the regions $|\arg k|<C$ may have a limiting point only at infinity.

3) The truncated resolvent $\widehat{R}_{k^2}$ has a pole at a real $k\neq 0$ (with $\arg k=0$ or $\pi$) if and only if the homogeneous problem (\ref{fk1}) has a non-trivial solution satisfying the radiation condition:
\[
|u|< Cr^{-(d-1)/2}, \quad |\frac{\partial u}{\partial r}-iku|<Cr^{-(d+1)/2}, \quad r=|x|\to\infty.
\]
\end{theorem}
{\bf Remarks.} 1) The first statement implies that the spectrum of $H_\beta$ outside of $[0,\infty)$ consists of a discrete set of eigenvalues of finite multiplicity with the only possible limiting point at infinity. While the last statement of the theorem indicates the possibility of the existence of spectral singularises on the continuous spectrum, see \cite{klv}, operator $H_\beta$ does not have eigenvalues imbedded into the continuous spectrum. The latter follows from the arguments used in \cite[Theorem 3.3]{66}.

2) The FKW problem has a non-local boundary condition, and therefore it is not elliptic. It is also non-symmetric, unless $\mu$ in (\ref{fk1}) is the Lebesgue measure. Theorem \ref{t41} is known \cite{75} for general (non-symmetric) exterior elliptic problems with fast stabilizing at infinity coefficients (see also \cite{vbook}). There is a wide literature concerning estimates on eigenvalues of non-symmetric elliptic problems, see for example \cite{davis,frank,BO,saf,wang,dem,fr} and references therein.  In particular, \cite{wang} contains the proof of the finiteness of the number of eigenvalues for the Schr\"{o}dinger operators with complex potentials in $R^d$ under certain assumptions on the potential with a minimal requirement on the decay rate at infinity. Note that a similar result is not valid for  exterior problems with fast decaying potentials, where the number of eigenvalues can be infinite even in the one-dimensional case \cite{pav}.

{\bf Proof.} As we mentioned above, the statement of the theorem is well known \cite{75} for the resolvent $R_{\lambda,D}$ (and the truncated resolvent $\widehat{R}_{\lambda,D}$) of the problem with the Dirichlet boundary condition (as well as for other elliptic boundary conditions). In particular, from \cite{75} it follows that the problem
\begin{equation}\label{fk3}
-{\rm div} (a(x)\nabla v) -\beta V(x)v-\lambda v=f, \quad x\in \Omega\subset R^d; \quad
v|_{\partial\Omega}=1, \quad \lambda\in \mathbb C\backslash [0,\infty),
\end{equation}
with $f\in H^{s}(\Omega_{com})$ has a meromprphic in $\lambda$ solution $v\in H^{s+2}(\Omega)$, and $\chi (x)v$ has a meromorphic continuation in $k=\sqrt \lambda$ with the properties described in the second statement of the theorem above. These properties of $v$ follow immediately from the properties of $\widehat{R}_{\lambda,D}$ after the substitution $v=\phi+w$, where $\phi\in C^\infty_0,~\phi=1$ in a neighborhood of $\partial\Omega$, and $w$ is the solution of the corresponding Dirichlet problem.

Let us look for the solution $u\in H^{s+2}(\Omega)$ of (\ref{fk1}) in the form
\begin{equation}\label{uv}
u=\alpha v+R_{\lambda,D}f, ~~~\lambda\in \mathbb C\backslash [0,\infty).
\end{equation}
Obviously, $u$ satisfies (\ref{fk1}) if and only if
\begin{equation}\label{al}
\alpha=-\gamma(\lambda,f)/\gamma_1(\lambda), \quad {\rm where} \quad \gamma(\lambda,f)=\int_{\partial\Omega}\frac{\partial}{\partial n}R_{\lambda,D}fd\mu,~~\gamma_1(\lambda)=\int_{\partial\Omega}\frac{\partial v}{\partial n}d\mu.
\end{equation}
From the analytic properties of $v$ it follows that $\gamma_1$ is meromorphic in $\lambda\in \mathbb C\backslash [0,\infty)$ and admits a meromorphic continuation to the whole complex $k$-plane if $d$ is odd or to the Riemann surface of $\ln k$ if $d$ is even. When $\lambda\in \mathbb C\backslash [0,\infty)$, function $v$ decays at infinity. If $\lambda<-\beta\max V(x)$, then the maximum principle is valid for solutions of (\ref{fk3}), $v$ achieves its maximum value at all the points of the boundary, and therefore, $\frac{\partial v}{\partial n}>0$ on $\partial\Omega$. Thus $\gamma_1(\lambda)>0$ when $\lambda<-\beta\max V(x)$. Hence $\gamma_1(\lambda)\not\equiv 0$, and therefore  $\gamma_1^{-1}(\lambda)$ is meromorphic in the complex $k$-plane if $d$ is odd or on the Riemann surface of $\ln k$ if $d$ is even. Moreover, from the  asymptotics of $\widehat{R}_{\lambda,D}$ as $\lambda\to 0$ \cite[Theorem 10]{75} it follows that the origin is not a limiting point for zeroes of $\gamma_1^{-1}(\lambda)$ located in a region $|\arg k|<C$. Thus, the poles of $\gamma_1^{-1}(\lambda)$ in this region may converge only to infinity, and therefore the first two statements of Theorem \ref{t41} follow from (\ref{uv}), (\ref{al}), and the validity of these statements for $R_{\lambda,D}$. The last statement of the theorem can be proved in the same way as a similar statement for $\widehat{R}_{\lambda,D}$ was proved in \cite{68}.

\qed

Denote by $A_\lambda$ the operator $\widehat{R}_\lambda$ for problem (\ref{fk1}) with $ \beta =0$ and  $\chi=\sqrt V$, i.e.,
\begin{equation}\label{alya}
A_\lambda=\sqrt V(x)(H_0-\lambda)^{-1}\sqrt V(x):H^s(\Omega)\to H^{s+2}(\Omega),
\end{equation}
where $H_0$ is defined by (\ref{fk1}).
\begin{lemma}\label{lale}
The following relations are valid for operator (\ref{alya}):
\[
\|A_\lambda\|\leq C<\infty \quad {\rm as} ~~\lambda\to 0^-~~ {\rm if } ~~d\geq 3; \quad  \lim _{\lambda\to 0^-}\|A_\lambda\|=\infty \quad {\rm if } ~~d=1, 2.
\]
\end{lemma}
{\bf Proof.} Below we assume that $\beta=0$.

  Let $d\geq 3$. For each $\rho <\infty,~\Omega_\rho=\Omega\bigcap |x|<\rho$, and $\lambda\to 0^{-}$, the solution $v\in H^{s+2}(\Omega)$ of  (\ref{fk3}) converges in $H^{s+2}(\Omega_\rho)$ to a decaying at infinity solution of  the same equation with $\beta=\lambda= 0$. Hence, the arguments used in the proof of Theorem \ref{t41} to show that $\gamma_1(\lambda)>0$ for $\lambda<-\beta \max V(x)$ remain valid when $\beta=0,~\lambda= 0^-$, i.e., $\gamma_1(0^-)>0$. Hence, the first statement of the lemma follows immediately from (\ref{uv}), (\ref{al}), and the boundedness of  $\widehat{R}_{\lambda,D}$ as $\lambda\to 0^-$.

  If $d=1$ or $2$, then $v$ converges in each $H^{s+2}(\Omega_\rho)$ to a constant (equal to one) as $\lambda\to 0^-$, and therefore $\gamma_1(0^-)=0$.
 Thus the second statement of the lemma will follow from  (\ref{uv}), (\ref{al}) if we show the existence of $f\in H^s$ such that
 \begin{equation}\label{fu}
 \gamma(0^-,f)>c>0  \quad {\rm and} \quad \|\widehat{R}_{0^-,D}f\|<\infty.
 \end{equation}
  To construct such an $f$, we consider an arbitrary $u\in H^{s+2}(\Omega_{com})$ with a compact support and such that $u|_{\partial\Omega}=0,~\frac{\partial u}{\partial n}|_{\partial\Omega}=1$. We have
 \[
H_0u-\lambda u=f-\lambda u, \quad {\rm where} ~~f=-{\rm div} (a(x)\nabla u)\in H^{s}(\Omega_{com}).
 \]
One can assume that the cut-off function $\chi$ in the definition of $\widehat{R}_{\lambda,D}$ is chosen in such a way that $\chi=1$ on the support of $u$. Then $\widehat{R}_{\lambda,D}f=u-\lambda\widehat{R}_{\lambda,D}u$. If $d=1,2$, then $\|\widehat{R}_{\lambda,D}u\|_{L^2}<C<\infty,~\lambda\to 0^-,$ due to Theorem \ref{tmr}. From a priori estimates for elliptic equations, it follows that the same estimate holds in the space $H^{s+2}.$ Hence $\|\widehat{R}_{\lambda,D}f-u\|_{H^{s+2}}\to 0$ as $\lambda\to 0^-$, and this implies (\ref{fu}).

\qed

As me mentioned earlier, Theorem \ref{tres} remains valid for symmetric FKW problems. Thus Lemma \ref{lale} implies the following statement.
\begin{theorem}
If $\mu$ is the Lebesgue measure on the boundary in FKW problem, then $\beta_{cr}=0$  in dimensions one and two and $\beta_{cr}>0$ if $d\geq 3$.
\end{theorem}

\bibliographystyle{amsalpha}

\end{document}